\documentclass[%
 reprint,
superscriptaddress,
 amsmath,amssymb,
 aps,
]{revtex4-1}

\usepackage{graphicx}
\usepackage{dcolumn}
\usepackage{bm}

\begin{document}

\newcommand{\mvec}[2]
{
\left(\begin{array}{c}
#1  \\
#2  
\end{array}
\right)
}

\newcommand{\mmat}[4]
{
\left(\begin{array}{cc}
#1  & #2\\
#3  & #4
\end{array}
\right)
}

\newcommand{\mvecthree}[3]
{
\left[
\begin{array}{c}
#1  \\
#2  \\
#3  
\end{array}
\right]
}
\newcommand{\defto}{\stackrel{\rightharpoonup}{=}}
\newcommand{\deffrom}{\stackrel{\leftharpoonup}{=}}
\newcommand{\mi}{{\mathrm i}}
\newcommand{\cmt}[2]{{[}#1,#2{]}}
\newcommand{\acmt}[2]{{\{}#1,#2{\}}}


\title{Lattice realization of the generalized chiral symmetry in two dimensions}

\author{Tohru Kawarabayashi}
\affiliation{Department of Physics, Toho University,
Funabashi, 274-8510 Japan}

\author{Hideo Aoki}
\affiliation{Department of Physics, University of Tokyo, Hongo, 
Tokyo 113-0033 Japan }
\affiliation{Electronics and Photonics Research Institute, 
Advanced Industrial Science and Technology (AIST), 
Umezono, Tsukuba, Ibaraki 305-8568, Japan}

\author{Yasuhiro Hatsugai}
\affiliation{Institute of Physics, University of Tsukuba, Tsukuba, 305-8571 Japan}

\date{\today}

\begin{abstract}
While it has been pointed out that the chiral symmetry, which is important 
for the Dirac fermions in graphene, can be generalized to tilted Dirac fermions 
as in  organic metals, such a generalized symmetry was so far defined only for a continuous low-energy Hamiltonian.
Here we show that the generalized chiral symmetry can be 
rigorously defined for lattice fermions as well. 
A key concept is a continuous ``algebraic deformation" of 
Hamiltonians, which generates 
lattice models with the generalized chiral symmetry from 
those with the conventional chiral symmetry. 
This enables us to explicitly express zero modes of the deformed Hamiltonian 
in terms of that of the original Hamiltonian.  
Another 
virtue is that the deformation  can be extended to non-uniform systems, such as
fermion-vortex systems and disordered systems. 
Application to fermion vortices in a deformed system shows how the zero modes 
for the conventional Dirac fermions with vortices 
can be extended to the tilted case.
\end{abstract}


\maketitle

\section{Introduction}

The chiral symmetry has served as one of the important symmetries in classifying the disordered systems \cite{AZ}. For the two-dimensional 
massless Dirac fermions as in graphene \cite{Geim,Kim,ZhengAndo}, it has been shown that the symmetry protects the zero-mode Landau levels, 
which gives rise 
to a criticality of the quantum Hall transition at the charge-neutrality point \cite{OGM,KHA1,KHA2,HA1}. The chiral symmetry has then been extended to encompass more 
general cases, namely tilted Dirac fermions such as 
observed in an organic compound $\alpha$-(BEDT-TTF)$_2$I$_3$ \cite{TSTNK,KKS,KKSF,KSFG,KNTSK,GFMP,MHT}, where 
the robustness of the zero modes is retained for tilted massless as well as 
massive Dirac fermions \cite{KHMA,HKA}. So far, however, 
the generalization has only been considered for
the Dirac field in low-energy, effective Hamiltonians, so that 
it remains unclear whether lattice fermions respecting the generalized chiral symmetry can be constructed or even exist. 
Here, we explore exactly this issue, and we shall 
show how such a generalized chiral 
symmetry can be extended to lattice fermions.  
This is not only conceptually interesting, but would also 
facilitate numerical analyses based on 
lattice models to clarify the effect of symmetry.  
A key idea here is an introduction of a ``continuous deformation" of Hamiltonians having the generalized chiral symmetry. The deformation, which does not change the basic profile of  the zero-energy state, 
can be applied not only to effective Hamiltonians in the continuum limit, but also to lattice models.  The deformation also turns out to be applicable to 
spatially non-uniform systems, so we shall discuss as a spin-off 
how the zero-energy solutions in the fermion-vortex system considered by 
Jackiw and Rossi \cite{JR,Weinberg} are generalized for tilted Dirac fermions. 

The conventional chiral symmetry is defined by the chiral operator $\Gamma$ (with $\Gamma^2 =1$) that anti-commutes with the
Hamiltonian $H$  ($\Gamma H\Gamma =-H$). For the conventional Dirac fermions 
as in graphene, the low-energy, effective Hamiltonian is expressed 
as $H_0 = v_F (\sigma_x p_x \pm \sigma_y p_y)$. The conventional chiral operator is then simply $\Gamma = \pm \sigma_z$. 
Here $(\sigma_x, \sigma_y, \sigma_z)$ are Pauli matrices and $(p_x,p_y)$ the momentum. 
The conventional chiral symmetry 
is defined not only for effective Hamiltonians, but also for lattice models with a bipartite structure. The two-dimensional honeycomb lattice for 
graphene is indeed typical, where the lattice can be divided into A and B sub-lattices with transfer integrals only between A and B.     
It is then straightforward to see that the chiral operator reads, 
on the lattice, $\Gamma = \exp(\mi\pi \sum_{n \in A} c_n^\dagger c_n)$, where $c_n^\dagger(c_n)$ denotes 
the creation (annihilation) operator of an electron at atomic site $n$. 
On the other hand, the generalized chiral symmetry has so far been defined only for the low-energy, effective Hamiltonian having tilted Dirac cones. 
Thus our goal is to explicitly 
construct or even generate systematically lattice models that have the rigorous generalized chiral symmetry.

Let us begin with the generalized chiral symmetry defined by the generalized chiral operator $\gamma$ ($\gamma^2 = 1$), 
which is not hermitian but satisfies $\gamma^\dagger H \gamma = -H$ \cite{KHMA,HKA}. For general tilted Dirac fermions described by the effective Hamiltonian, 
$$
 H = [-X_0 +(\bm{X}\cdot \bm{\sigma})]p_x +  [-Y_0 +(\bm{Y}\cdot \bm{\sigma})]p_y,
$$
with $\bm{X}$ and $\bm{Y}$ being three-dimensional real vectors, such $\gamma$ exists as long as a condition, $(\bm{X}\times\bm{Y})^2 - \bm{\eta}^2 >0 $, is fullfilled with $\bm{\eta}= Y_0 \bm{X} -X_0 \bm{Y}$, which is 
equivalent to the ellipticity of the Hamiltonian $H$ as a differential operator \cite{Nakahara}. 
An explicit expression for the generalized chiral operator $\gamma$ is
\begin{eqnarray}
 \gamma = T_{\bm{\tau}_0} \  \Gamma  \ T_{\bm{\tau}_0}^{-1} ,\label{gcop}\\
T_{\bm{\tau}_0} = \exp(q \bm{\tau}_0\cdot \bm{\sigma} /2),\nonumber
\end{eqnarray}
where $\bm{\tau}_0$ is a unit vector parallel to $(\bm{X}\times\bm{Y})\times \bm{\eta}$ 
and $\tanh q = |\bm{\eta}|/|\bm{X}\times \bm{Y}|$. The conventional chiral operator $\Gamma$ for the vertical Dirac fermion ($X_0=Y_0=0$) is 
given by $\Gamma = (\bm{X}\times \bm{Y})\cdot \bm{\sigma}/|\bm{X}\times \bm{Y}| $. Note that the operator $T_{\bm{\tau}_0}$ is not
unitary but hermitian ($T_{\bm{\tau}_0}^\dagger = T_{\bm{\tau}_0}$) with $q$ real as long as $|\bm{X} \times \bm{Y}|>|\bm{\eta}|$.
We shall use the algebraic expression (\ref{gcop}) for the generalized chiral operator $\gamma$ in terms of the conventional chiral operator $\Gamma$ 
to propose a systematic deformation of Hamiltonians preserving the generalized chiral symmetry.
We actually consider an algebraic deformation $H_{\bm{\tau}} = T_{\bm{\tau}}^{-1}H_0T_{\bm{\tau}}^{-1}$ of the original lattice Hamiltonian 
$H_0$, respecting the conventional chiral symmetry for 
 vertical Dirac fermions, using a hermitian matrix $T_{\bm{\tau}}$. We shall 
show that the deformed lattice Hamiltonian $H_{\bm{\tau}}$ exactly respects 
the generalized chiral symmetry and hosts the tilted Dirac fermions in two dimensions.

The present paper is organized as follows. After the introduction 
in section 2, we extend  in section 3 our deformation 
to lattice fermions, where Dirac fermions are always doubled. In section 4. we analyze the consequences of our 
deformation in lattice models with the translational invariance. An application to the  zero modes of the fermion-vortex system where 
the translational invariance is broken, is given in section 5. Section 6 is devoted to summary.

\section{$q$-deformation for single\\ 
Dirac fermion}

Before a full description of the general deformation for lattice fermions, 
let us first discuss, for illustrative purpose, a 
deformation for effective, single Dirac fermions. We 
define a deformation $H_{\bm{\tau}}(q)$ of the original effective Hamiltonian $H_0$  as
\begin{equation}
 H_{\bm{\tau}}(q) = [T_{\bm{\tau}}(q)]^{-1} \ H_0 \  [T_{\bm{\tau}}(q)]^{-1} 
 \label{q-def}
\end{equation}
with 
$$
 T_{\bm{\tau}}(q) = \exp\bigg(\frac{q}{2}\bm{\tau}\cdot\bm{\sigma} \bigg),
$$
where $\bm{\tau}$ denotes an {\it arbitrary} unit vector and $q$ a real parameter,
which we call  ``$q$-deformation" in the following.
Note that $H_{\bm{\tau}}(q)$ is hermitian, since $T_{\bm{\tau}}(q) $ is.
We assume that the original Hamiltonian $H_0$ respects the conventional chiral symmetry, so that there exists a
chiral operator $\Gamma$ satisfying $\Gamma H_0 \Gamma = -H_0$ with $\Gamma^2=1$.
We then define a generalized chiral operator $\gamma$ by 
\begin{equation}
 \gamma = T_{\bm{\tau}}(q) \ \Gamma \ [T_{\bm{\tau}}(q)]^{-1}.
 \label{ch-op}
\end{equation}
It is  straightforward to see that 
\begin{eqnarray}
 \gamma^\dagger H_{\bm{\tau}}(q) \gamma &=& [T_{\bm{\tau}}(q)]^{-1} \Gamma H_0 \Gamma [T_{\bm{\tau}}(q)]^{-1} = -H_{\bm{\tau}}(q) ,
 \label{gcs}\\
\gamma^2&=&1. \nonumber
\end{eqnarray}
The  $q$-deformation (\ref{q-def}) 
therefore generates systems with the exact generalized chiral symmetry from those with the conventional chiral symmetry.

One of the important properties of this deformation is that the wave function of zero modes are explicitly given in terms of those of the  
original Hamiltonian $H_0$. If we have a zero mode $\psi_0$ of the original $H_0$ satisfying $H_0\psi_0 =0$, the corresponding zero-mode of the deformed 
Hamiltonian is given by a simple transformation
$T_{\bm{\tau}}(q) \psi_0$, which is non-unitary \cite{HT,LSB,SGT},  since $H_{\bm{\tau}}(q) [T_{\bm{\tau}}(q) \psi_0] = T_{\bm{\tau}}^{-1}(q)H_0\psi_0 =0$.
The zero modes are thus retained by this deformation. Furthermore, if we recall that the zero modes of the original Hamiltonian can be taken as 
the eigenstates of the chiral operator $\Gamma$ as $\Gamma \psi_0 = \pm \psi_0$ \cite{KHA1}, 
the transformed zero modes $T_{\bm{\tau}}(q) \psi_0$ of the deformed Hamiltonian 
become the exact eigenstates of the generalized chiral operator $\gamma$ as 
$\gamma\  T_{\bm{\tau}}(q) \psi_0 = T_{\bm{\tau}}(q) \Gamma \psi_0 = \pm T_{\bm{\tau}}(q) \psi_0 $. 
We can also note that the determinant of the 
Hamiltonian is invariant  in the deformation ($\det H_q = \det H_0$), which follows from $\det T_{\bm{\tau}}(q)= 1$.

It is verified directly that the present deformation indeed produces tilted Dirac fermions from vertical Dirac fermions. Namely, 
from $H_0 = (\bm{X}\cdot \bm{\sigma})p_x +  (\bm{Y}\cdot \bm{\sigma})p_y$ for vertical Dirac fermions, we obtain 
\begin{eqnarray*}
 H_{\bm{\tau}}(q)  &=& (-X_0'+\bm{X}'\cdot \bm{\sigma}) p_x 
  +(-Y_0' + \bm{Y}'\cdot \bm{\sigma}) p_y 
\end{eqnarray*}
with 
\begin{eqnarray*}
X_0' &=&  \sinh q (\bm{\tau}\cdot\bm{X}), \ Y_0' =  \sinh q (\bm{\tau}\cdot\bm{Y}),\\
 \bm{X}' &=& \bm{X} +(\cosh q -1)(\bm{\tau}\cdot\bm{X})\bm{\tau}, \\
  \bm{Y}'&=& \bm{Y} +(\cosh q -1)(\bm{\tau}\cdot\bm{Y})\bm{\tau}.
\end{eqnarray*}
This is nothing but the Hamiltonian for the tilted Dirac fermions except for the case $\bm{\tau} \propto (\bm{X}\times \bm{Y})$.

Here the vector $\bm{\tau}$, which defines the present deformation, can be chosen arbitrarily, and is in  principle independent of 
the vector $\bm{n}=\bm{X}\times \bm{Y}/|\bm{X}\times \bm{Y}|$ characterizing the conventional chiral operator, $\Gamma = \bm{n}\cdot \bm{\sigma}$. 
However, we can emphasize that, 
even when the vector $\bm{\tau}$ has a component parallel to $\bm{n}$, such a component does not contribute to the tilting of the Dirac fermions, hence 
to the breaking of the conventional chiral symmetry.  
If we resolve $\bm{\tau}$ into the components parallel $\bm{\tau}_\parallel$ and perpendicular $\bm{\tau}_\perp$ to $\bm{n}$, we see that the parameters $(X_0',Y_0')$ are indeed determined only by $\bm{\tau}_\perp$, since 
$(\bm{\tau}\cdot \bm{X}) = 
(\bm{\tau}_\perp\cdot \bm{X})$ and  $(\bm{\tau}\cdot \bm{Y}) = 
(\bm{\tau}_\perp\cdot \bm{Y})$.
In the present paper,  we therefore focus ourselves on the deformation where $\bm{\tau}$ is perpendicular to $\bm{n}$.

\section{Generalization to Lattice fermions}
\subsection{General formalism}

Now we show that the $q$-deformation can be extended to lattice models. 
Generally, the Hamiltonian of a lattice model that can be reduced to a form, 
\begin{eqnarray}
H &=&  \mmat{D_z}{D_x-\mi D_y}{D_x+\mi D_y}{-D_z}  \label{general_H} \\  
 &=& \bm{D}\cdot\bm{\sigma} \nonumber
\end{eqnarray}
with $D_x$, $D_y$ and $D_z$ being $N \times N$ hermitian matrices, 
has the chiral symmetry if there exits a real vector $\bm{n} = (n_x, n_y, n_z)$ with $\bm{n}^2 = 1$ satisfying the condition  
$$
 \bm{D}\cdot \bm{n} =0.
$$ 
This can be verified with the chiral operator defined as 
$$
 \Gamma = I_N \otimes \bm{n}\cdot \bm{\sigma}
$$
that anti-commutes with the Hamiltonian, $ \Gamma H \Gamma = -H$ and $\Gamma^2 = I_{2N}$, 
where $I_N$ denotes the $N\times N$ identity matrix.

To be more specific, we consider  non-interacting fermions on a  lattice with a bipartite structure respecting 
the conventional chiral symmetry. Bipartite lattice models  can 
be expressed as 
$$
H_{\rm c} =  \mmat{O}{D_x-iD_y}{D_x+iD_y}{O} = D_x \otimes \sigma_x + D_y \otimes \sigma_y
$$
in a basis $(a_1, \ldots , a_N, b_1, \ldots , b_N)$, where $a_i(b_i)$ denotes
the basis on the A(B) sub-lattice in the $i$th unit cell. Here we assume that 
the two sub-lattices have the same number of sites $N$, 
while the case of different numbers of sub-lattice sites is 
considered in Appendix B.
The conventional chiral operator can then be defined as 
$$
 \Gamma = \mmat{I_N}{O}{O}{-I_N}  = I_N \otimes \sigma_z
$$  
by setting $\bm{n}=(0,0,1)$.

We then define the $q$-deformation of such a chiral symmetric Hamiltonian 
as 
$$
 H_{\bm{\tau}}(q) = [T_{\bm{\tau}}(q)]^{-1} \ H_{\rm c} \ [T_{\bm{\tau}}(q)]^{-1}
$$  
with 
$$
 T_{\bm{\tau}}(q) = I_N \otimes \exp\bigg(\frac{q}{2}\bm{\tau}\cdot\bm{\sigma}\bigg),
$$
where the 
generalized chiral operator is
$$
 \gamma = T_{\bm{\tau}}(q)\  \Gamma \ [T_{\bm{\tau}}(q)]^{-1}.
$$ 
Here $\bm{\tau}=(\tau_x,\tau_y,\tau_z)$ is an arbitrary three-dimensional real vector with the unit length ($\bm{\tau}^2 =1$) and $q$ a real parameter.  
We can readily see that the deformed lattice Hamiltonian $H_{\bm{\tau}}(q)$ respects the 
generalized chiral symmetry, $\gamma^\dagger H_{\bm{\tau}}(q) \gamma = -H_{\bm{\tau}}(q)$.
It should be noted  that this deformation can be performed for a wide variety  of systems with or without the 
translational invariance, which cover disordered systems and 
fermion-vortex systems.

In this representation, the deformation becomes non-trivial only for the cases $\tau_x\neq 0$ or $\tau_y\neq 0$. 
As discussed toward the end of Section 2, we assume $\bm{\tau}\cdot \bm{n}=0$, namely $\tau_z=0$, for the deformation of bipartite lattice models.
A non-trivial deformation respecting the time-reversal symmetry is therefore possible only  when $\tau_x \neq 0$ and $\tau_y =0$, for which the matrix 
$T_{\bm{\tau}}(q) = I_N \otimes \exp(q\sigma_x/2)$ becomes  real and symmetric (see Appendix A). By contrast, a deformation with $\tau_y\neq 0$,e.g. $T_{\hat{\bm{y}}}(q) = 
\exp(q\sigma_y/2)$, would break the time-reversal invariance because the matrix $T_{\hat{\bm{y}}}(q)$ has complex matrix elements that induce 
additional complex transfer integrals of the deformed lattice Hamiltonians.

Since we can assume $\tau_z=0$, we can represent $\bm{\tau} = (\cos \theta, \sin \theta ,0)$. The matrix $T_{\bm{\tau}}(q)$ is then
given by
\begin{eqnarray*}
  T_{\bm{\tau}}(q) &=& 
  =
  I_N \otimes \mmat{\cosh \frac{q}{2} }     {e^{-\mi \theta }\sinh \frac {q}{2} }
      {e^{\mi \theta }\sinh \frac {q}{2} }      {\cosh \frac {q}{2} }.
\end{eqnarray*} 
Note that we have a relation $\Gamma\  T_{\bm{\tau}}(q) = T_{\bm{\tau}}(-q) \ \Gamma $.
The deformed Hamiltonian  
$H_{\bm{\tau}}(q) = T_{\bm{\tau}}(q)^{-1} \ H_{\rm c} \ T_{\bm{\tau}}(q)^{-1}$  then becomes 
$$
 H_{\bm{\tau}}(q) = -\frac{\sinh q}{2} (e^{\mi \theta }D+e^{-\mi \theta }D^\dagger)   \otimes I_2 + \mmat{O}{D_q}{D_q^\dagger}{O},
$$
where 
$$
 D_q = D\cosh^2 \frac{q}{2} +e^{-2\mi \theta }D ^\dagger \sinh ^2 \frac {q}{2} 
$$
with $D \equiv D_x-iD_y$.
If we further define
\begin{eqnarray*}
    \bar H_{\bm{\tau}}(q) &\equiv &  H_{\bm{\tau}}(-q)\\
    &=& 
H_{\bm{\tau}}(q) +  (e^{\mi \theta }D +e^{-\mi \theta }D^\dagger) \sinh q \otimes I_2,
 \end{eqnarray*}
we have a simple relation, 
\begin{eqnarray*}
    \bar H_{\bm{\tau}}(q) H_{\bm{\tau}}(q) &=&
    T_{\bm{\tau}}(q)  H_{\rm c} ^2 [T_{\bm{\tau}}(q)]^{-1}    ,
 \end{eqnarray*}
which can be compared with the counterpart in the continuum \cite{HKA}.
If we denote the eigenstate of the Hamiltonian $H_{\bm{\tau}}(q)$ with an eigenenergy $E$ as $\psi_E$,
this relation implies 
\begin{eqnarray*}
\lefteqn{\bar H_{\bm{\tau}}(q) H_{\bm{\tau}}(q) \psi_E} \quad\\
   & &= \left[E^2 I_{2N} + E(e^{\mi \theta }D+ e^{-\mi \theta }D^\dagger) \sinh q \otimes I_2\right]\psi_E\\
    & &= 
   T_{\bm{\tau}}(q) H_{\rm c}^2 T_{\bm{\tau}}(q)^{-1}\psi_E ,
\end{eqnarray*}
which leads to 
\begin{eqnarray*}
T_{\bm{\tau}}(q) \big[ H_{\rm c}^2  -E(e^{\mi \theta }D +e^{-\mi \theta }D^\dagger) \sinh q \otimes I_2\big]T_{\bm{\tau}}(q)^{-1}\psi_E & & \\
   = E^2\psi_E & &.
\end{eqnarray*}
Completing the square as 
  \begin{eqnarray*}
&&    H_{\rm c}^2 -(e^{\mi \theta }D +e^{-\mi \theta }D^\dagger) E\sinh q \otimes I_2 
\\    &=&
    \mmat{O}{D}{D ^\dagger }{O} ^2
    -(e^{\mi \theta } D+e^{-\mi \theta } D ^\dagger ) E\sinh q \otimes I_2 
    \\
    &=& 
    \mmat{O}{D(E)}{D(E)^\dagger}{O} ^2
    - E^2 \sinh^2 q \otimes I_{2N}, 
  \end{eqnarray*}
where $D(E) \equiv D-Ee^{-\mi \theta }\sinh q$ and we have defined a nonorthogonal state $\Psi$ with an ``overlap'' matrix $T_{\bm{\tau}}(2q)$ as
  \begin{eqnarray*} 
    \Psi &=& T_{\bm{\tau}}(q) ^{-1}   \psi_E,\ \
    \Psi ^\dagger T_{\bm{\tau}}(2q)\Psi = \psi_E^\dagger \psi_E=1,
  \end{eqnarray*}
  we arrive at  an  eigenvalue problem, 
  \begin{eqnarray*}
    [H_q ^R(E)]^2\Psi &=& E^2\Psi,
  \end{eqnarray*}
  where the Hamiltonian $H_q^R(E)$ includes $E$ as a parameter as
  \begin{eqnarray*} 
    H_q^R (E)
    &=&  \mmat{O}{D_R(E) }{D_R ^\dagger (E)}{O}
    ,
\\
    D_R (E)&\equiv&  \frac {1}{\cosh q} (D- E e^{-\mi \theta } \sinh q).
  \end{eqnarray*}

\subsection{Chiral symmetry breaking for lattice fermions}

We can utilize the above  formulation for discussing the effect of the symmetry breaking when a mass term, $m\Gamma$, is introduced.
In the presence of the mass term, the $q$-deformed Hamiltonian becomes 
$$
 H_{\bm{\tau}}^{(m)}(q) = T_{\bm{\tau}}(q)^{-1} (H_{\rm c}+m\Gamma)T_{\bm{\tau}}(q)^{-1} = H_{\bm{\tau}}(q) + m\Gamma.
$$
If we define $\bar H_{\bm{\tau}}^{(m)}(q) = \bar H_{\bm{\tau}}(q) +m\Gamma$, we have 
\begin{eqnarray*}
    \bar H_{\bm{\tau}}^{(m)}(q) H_{\bm{\tau}}^{(m)}(q) &=&
\bar H_{\bm{\tau}}(q) H_{\bm{\tau}}(q)+ m^2\\
 & = & T_{\bm{\tau}}(q) (H_{\rm c}^2 +m^2)T_{\bm{\tau}}(q)^{-1} .
\end{eqnarray*}
Multiplying these operators to an eigenstate $\psi^{(m)}_E$ of $H_{\bm{\tau}}(q)$,
we find
\begin{eqnarray*}
\lefteqn{\bar H_{\bm{\tau}}^{(m)}(q) H_{\bm{\tau}}^{(m)}(q)\psi^{(m)}_E}\\
&=&  [E^2 + E(e^{\mi \theta }D+e^{-\mi \theta }D^\dagger)  \sinh q +m^2] \psi^{(m)}_E\\
&=&     T_{\bm{\tau}}(q) (H_{\rm c}^2 +m^2)T_{\bm{\tau}}(q) ^{-1} \psi^{(m)}_E,
\end{eqnarray*}
which gives 
\begin{eqnarray*}
  \big[H_{\rm c}^2 -(e^{\mi \theta }D+e^{-\mi \theta }D^\dagger) E\sinh q +m^2\big]\Psi^m &=& E^2\Psi^m
\end{eqnarray*}
with $\Psi^m=T_{\bm{\tau}}(q) ^{-1}\psi^{(m)}_E$. 
We finally arrive at 
\begin{eqnarray}
\left\{[ (H_q^R(E)]^2 + m_R^2\right\}\Psi^m = E^2\Psi^m, \label{massiveeq}
  \\
  m_R \equiv \frac {m}{\cosh q}.\nonumber
\end{eqnarray} 
This means that $E^2$ is generally lower-bounded by $m_R^2$.
Furthermore, if we have a zero mode for $ H_q^R(m_R)$ with
$$
 H_q^R(m_R) \Psi^m = 0,
$$ 
the eigenvalue $E$ of Eq.(\ref{massiveeq}) is exactly $\pm m_R$ with 
an energy gap $2m_R$. The sign of $E$ can be determined to be consistent with the sign in the limit $q \rightarrow 0$.
If we assume that the eigenstate is that of the chiral operator $\Gamma$ with 
an eigenvalue $+1$ as 
$$
 \Psi^m = \left( \begin{array}{c} 
                                            \varphi_+^m\\
                                            0
                        \end{array}\right),
$$ 
then the equation becomes 
$$
 D_R (m_R)^\dagger \varphi_+^m=  \frac {1}{\cosh q} (D^\dagger- m_R e^{\mi \theta } \sinh q)\varphi_+^m=0.
$$
Note that in this case the energy must be positive, since it must approach to $+m$ for $q \to 0$. 

\begin{figure}[h]
\includegraphics[scale=0.35]{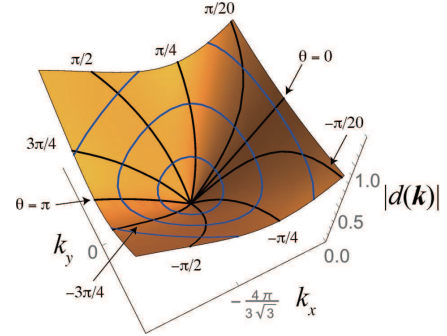}
\caption{(Color Online)
The absolute value $|d(\bm{k})|$ of the  complex number $d(\bm{k}) = |d(\bm{k})| e^{\mi \theta_d}$ for the honeycomb lattice around valley K, where 
$\bm{k}_0=(-\frac{4\pi}{3\sqrt{3}a},0)$ with $a$ bond length. The curves starting from $\bm{k}_0$ (center of the figure) represent 
phases $\theta_d = 0$, $\pm \pi/20$, $\pm \pi/4$, $\pm \pi/2$, $ \pm 3\pi/4$ ,  $\pi$. 
The phase $\theta_d$ changes from $0$ to $2\pi$ as we go around the singular point $\bm{k}_0$.  
\label{fig0}
}
\end{figure}

For a translationally invariant system, this reduces for each momentum sector specified by a wave vector $\bm{k}$ to 
\begin{equation}
  d^*(\bm{k})- m_R e^{\mi \theta } \sinh q=0,
  \label{app1}
\end{equation}
where $d(\bm{k})$ is a complex number which appears in the Hamiltonian in the momentum space as
\begin{equation}
 H_{\bm{k}} = \mmat{0}{d(\bm{k})}{d^*(\bm{k})}{0}.
 \label{Ham_k}
\end{equation}
The existence of doubled Dirac fermions at $\pm \bm{k}_0$ means that $d(\pm \bm{ k}_0)=0$ and the phases of the 
complex number $d(\bm{k})$ becomes indefinite at those points (Fig.\ \ref{fig0}). 
Assuming the continuity of a complex number $d(\bm{k})$ around the $\pm \bm{k}_0$, 
we may choose  $\bm{k}$ to 
satisfy the above equation (\ref{app1})  provided that the amplitude $m_R\sinh q$ is small enough.

Similarly, for an eigenstate in the form
$$
  \Psi^m = \left( \begin{array}{c} 
                                            0\\
                                            \varphi_-^m
                        \end{array}\right),
$$
we have  
the exact eigenvalue $-m_R$, if $\varphi_-^m$ satisfies the equation
$$
 D_R (-m_R) \varphi_-^m=  \frac {1}{\cosh q} (D+ m_R e^{-\mi \theta } \sinh q)\varphi_-^m=0.
$$

Another important observation  is that the exact zero modes of the deformed Hamiltonian, which is an eigenstate of the generalized chiral operator, 
has the energy expectation value $\pm m_R$ in the presence of the mass term $m\Gamma$ even for systems 
without translational invariance.
Let us assume that there exist a zero mode $\psi_+^c$ for an original Hamiltonian $H_{\rm c}$
satisfying 
$$
 D ^\dagger \phi_+^c = 0,
$$
where
\begin{eqnarray*}
    \psi_+^c &=& \mvec{\phi_+^c}{0} \ {\rm and}\ 
  H_{\rm c}= \mmat{O}{D}{D ^\dagger }{O}
\end{eqnarray*}
with a normalization $(\psi_+^c) ^\dagger \psi_+^c = 1$.
Now we define a state,
\begin{eqnarray*}
  \psi_+^m &=& \frac{1}{\sqrt{\cosh q}}T_{\bm{\tau}}(q) \psi_+^c, 
\end{eqnarray*}
which is an exact zero modes of the deformed Hamiltonian $H_{\bm{\tau}}(q)$.
We then have an expectation value, 
\begin{eqnarray*} 
 (\psi_+^m) ^\dagger [H_{\bm{\tau}}(q)+m \Gamma ]\psi_+^m    &=&  m/\cosh q = m_R,
\end{eqnarray*} 
since $H_{\bm{\tau}}(q) \psi_+^m=0$ and $m(\psi_+^m)^\dagger \Gamma \psi_+^m = +m_R$.
The expectation value of the Hamiltonian for the state $\psi_+^m$ gives the same value as the lower bound 
of the positive energy.

Similarly for the state
\begin{eqnarray*}
      \psi_-^{-m} &=& \frac{1}{\cosh q}T_{\bm{\tau}}(q) \psi_-^c,
\end{eqnarray*}
with
$$
      \psi_-^c = \mvec{0}{\phi_-^c}, 
      \ \
      D\phi_-^c = 0,
$$
we have 
\begin{eqnarray*} 
    (\psi_-^m) ^\dagger [H_{\bm{\tau}}(q)+m \Gamma ]\psi_-^m    &=&  -m/\cosh q = -m_R,
\end{eqnarray*}
because $m(\psi_-^m)^\dagger \Gamma \psi_-^m = -m_R$, which coincides with the upper bound of 
the negative energy.

\section{translationally invariant systems}

For systems with the translational invariance, the Hamiltonian can be reduced, in the momentum space, to a form 
(\ref{Ham_k})
and the deformed Hamiltonian given  as
$$
 H_{\bm{\tau}}(q) = \exp\left(-\frac{q}{2}\bm{\tau}\cdot\bm{\sigma}\right) H_{\bm{k}} \exp\left(-\frac{q}{2}\bm{\tau}\cdot\bm{\sigma}\right)
$$
with $\bm{\tau}=(\cos\theta, \sin \theta,0)$ then becomes  
 \begin{eqnarray*}
 H_{\bm{\tau}}(q)&=&-\sinh q {\rm Re}(e^{\mi \theta}d(\bm{k}))\\
  &+& \bigg[\cosh^2\frac{q}{2} {\rm Re}(d(\bm{k})) +\sinh^2\frac{q}{2} {\rm Re}(e^{\mi 2\theta}d(\bm{k}))\bigg]\sigma_x\\
  &-& \bigg[\cosh^2\frac{q}{2} {\rm Im}(d(\bm{k})) -\sinh^2\frac{q}{2} {\rm Im}(e^{\mi 2\theta}d(\bm{k}))\bigg]\sigma_y.
\end{eqnarray*}

In the following, we consider two examples of translationally invariant bipartite lattice models, namely, honeycomb lattice 
and the $\pi$-flux model on the square lattice.

\subsection{honeycomb lattice}

\begin{figure}[h]
\includegraphics[scale=0.57]{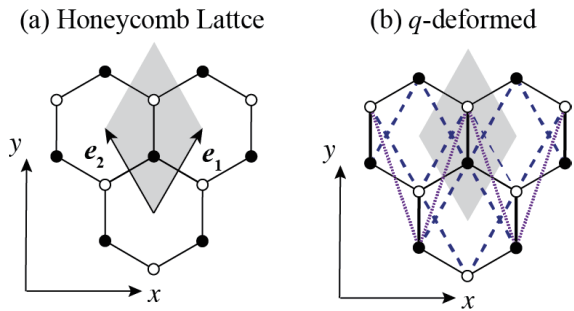}
\caption{(Color online)
(a) Honeycomb lattice, for which a unit cell (shaded) and the 
primitive vectors  $\bm{e}_1=(\sqrt{3}/2, 3/2)a$ and $\bm{e}_2=(-\sqrt{3}/2,3/2)a$ are indicated. The filled (open) circles represent A(B) sub-lattice sites. 
(b) The $q$-deformed honeycomb lattice model with $T_{\hat{\bm{x}}}(q)$, 
where second-neighbor (NN) hopping $-(t/2)\sinh q$ (dashed lines) and 
the 4th-neighbor hopping $(t/2)(\cosh q -1)$ 
(dotted) are generated. 
The NN hoppings are also modified to $(t/2)(\cosh q +1)$ (thin solid lines) 
across unit cells and 
$t\cosh q$ (thick solid lines) within a unit cell. 
The potential energies are  
modified uniformly to $-t\sinh q$.
\label{honeycomb}
}
\end{figure}

For the honeycomb lattice having only the nearest-neighbor hopping $t$, we have \cite{HFA}
$$
 d(\bm{k}) = t[1+\exp(-\mi k_1) + \exp(-\mi k_2)],
$$ 
where $(k_1,k_2)$ denotes $(\bm{k}\cdot\bm{e}_1, \bm{k}\cdot\bm{e}_2)$ with the primitive vectors $( \bm{e}_1,\bm{e}_2)$ of 
the honeycomb lattice (Fig. \ref{honeycomb}) defined as $\bm{e}_1 =( \sqrt{3}/2 \bm{e}_x + 3/2 \bm{e}_y)a$ and
$\bm{e}_2 = (-\sqrt{3}/2 \bm{e}_x + 3/2 \bm{e}_y)a$. Here $\bm{e}_{x(y)}$ stands for the unit vector along $x(y)$ and $a$ 
the nearest-neighbor distance of the honeycomb lattice. 

\begin{figure}[h]
\includegraphics[scale=0.5]{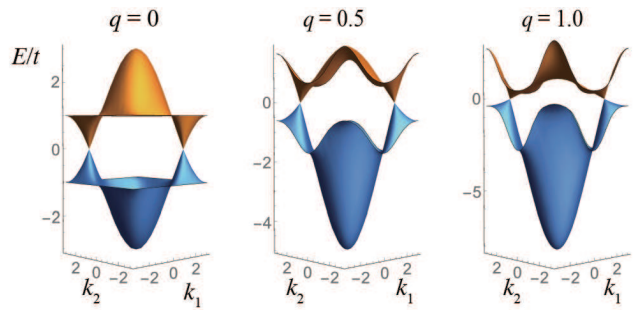}
\caption{(Color online)
Energy dispersions of the honeycomb lattice deformed  
by $T_{\hat{\bm{x}}}(q)$ with $q=0$ (left panel), $0.5$ (center) 
and $1.0$ (right).
Dirac cones at K and K' points are tilted along the $+k_x$ and $-k_x$ directions, respectively. 
\label{honeycomb_def}
}
\end{figure}

First, we consider the $q$-deformation respecting the time-reversal invariance, where $\bm{\tau} =\hat{\bm{x}}= (1,0,0)$, namely $\theta =0$.  
The $q$-deformed Hamiltonian then becomes 
$$
H_{\hat{\bm{x}}}(q) =  -\sinh q\ {\rm Re} (d) I_2 +\cosh q\ {\rm Re}(d) \sigma_x -{\rm Im}(d) \sigma_y,
$$
where $I_2$ is the $2\times 2$ identity matrix. The corresponding hoppings are 
displayed in Fig. \ref{honeycomb} (b).  
The energy dispersion  becomes 
\begin{eqnarray*}
 E_q(k_1,k_2) &=& -{\rm Re} (d)\sinh q \\
 & &\quad \pm \sqrt{({\rm Re}(d))^2\cosh^2 q + ({\rm Im}(d))^2},
\end{eqnarray*}
where the symmetry $E_q(k_1,k_2) = E_q(-k_1,-k_2)$ required by the time-reversal invariance (Appendix A) is satisfied.  
In this representation, K and K' points are at $(k_1,k_2) =(-2\pi/3,+2\pi/3)$ and $(2\pi/3,-2\pi/3)$, respectively.
If we expand the Hamiltonian around the K point, we have
$$
 H_{\rm K}^{\theta =0} = \frac{3ta}{2\hbar} (-\sinh q\ p_x I_2 + \cosh q\  p_x \sigma_x - p_y \sigma_y),
$$
where an  effective momentum $\bm{p}=\hbar \delta \bm{k}$ is 
defined in terms of $\delta \bm{k} = \bm{k} - \bm{k}_0$ with $\bm{k}_0$ being 
the wave vector at the K point.  
Similarly, we can derive the effective Hamiltonian for valley K' (Table \ref{tab1}).
We see that the isotropic and vertical Dirac fermions at the valleys K and K' of the honeycomb lattice are 
deformed into anisotropic and tilted Dirac fermions (Table \ref{tab1} and Fig.\ \ref{honeycomb_def}). Note that the tilting directions are opposite in the two valleys K and K' 
due to the time-reversal invariance.

\begin{table}
\begin{tabular}{c|cccc}
 & $X_0$ & $Y_0$ & $\bm{X}$ & $\bm{Y}$\\
\hline \hline
$H_{\hat{\bm{x}}}(q)$ at K & $\sinh q$ &0& $(\cosh q,0,0)$ & $(0,-1,0)$ \\
$H_{\hat{\bm{x}}}(q)$ at K' & $-\sinh q$ &0& $(-\cosh q,0,0)$ & $(0,-1,0)$\\
$H_{\hat{\bm{y}}}(q)$ at K & $0$ &$-\sinh q$& $(1,0,0)$ & $(0,-\cosh q,0)$ \\
$H_{\hat{\bm{y}}}(q)$ at K' & $0$ &$-\sinh q$& $(-1,0,0)$ & $(0,-\cosh q,0)$\\
\hline
\end{tabular}
\caption{For the honeycomb lattice the parameters $(X_0,Y_0,\bm{X},\bm{Y})$ normalized by $3ta/2\hbar$ are given 
for the effective massless Dirac Hamiltonian $(-X_0+\bm{X}\cdot\bm{\sigma})p_x +(-Y_0+\bm{Y}\cdot\bm{\sigma})p_y$ at valleys K and K'. For a general case where $\bm{\tau}$ is given by $(\cos \theta ,\sin \theta,0)
$, we have $X_0=\sinh q\cos \theta$, $Y_0=-\sinh q \sin \theta$, $\bm{X}=(\cosh^2\frac{q}{2}+\sinh^2\frac{q}{2}\cos2\theta,\sinh^2\frac{q}{2}\sin2\theta,0)$, $\bm{Y}=(-\sinh^2\frac{q}{2}\sin2\theta,-\cosh^2\frac{q}{2}+\sinh^2\frac{q}{2}\cos2\theta,0)$ for valley K. The parameters for valley K' are obtained 
by reversing the sign of $X_0$ and $\bm{X}$.}
\label{tab1}
\end{table}

In the present system, the staggered potential plays a role of the mass term, 
$m\sigma_z$. If we include this term, the energy dispersion is modified to
\begin{eqnarray*}
 E_q^m(k_1,k_2) &=& -{\rm Re} (d)\sinh q \\
 & &\quad \pm \sqrt{({\rm Re}(d))^2\cosh^2 q + ({\rm Im}(d))^2+m^2}.
\end{eqnarray*}
As discussed in Section 3, this can be rewritten in a form
$$
 d_R^*(E_q)d_R(E_q) +m_R^2 = E_q^2
$$
with $d_R(E_q) = [d(\bm{k}) - E_q\sinh q]/\cosh q$. The energy gap is therefore
given exactly as $\pm m_R=\pm m/\cosh q$ as long as we have a solution for $d(\bm{k}) - m_R\sinh q=0$, which is guaranteed for $m\tanh q \leq 3t$.

\begin{figure}[h]
\includegraphics[scale=0.56]{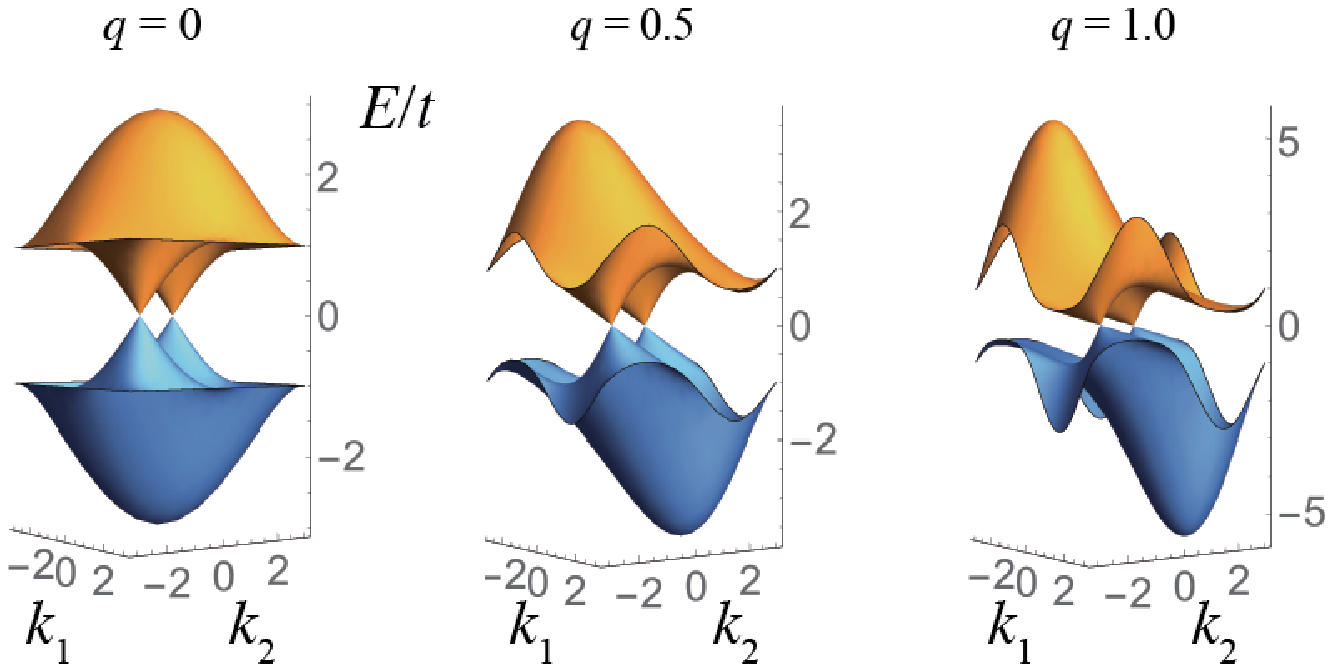}
\caption{(Color online)
Energy dispersions of the honeycomb lattice deformed by $T$-breaking 
$T_{\bm{y}}(q)$ with $q=0$ (left panel), $0.5$ (center) and $1.0$ (right). 
Dirac cones at K and K' points are both tilted along the $-k_y$ direction.
\label{honeycomb_def_y}
}
\end{figure}

Next, we consider the case $\theta =\pi/2$, where 
the deformation operator $T_{\hat{\bm{y}}}(q) = \exp(q\sigma_y/2)$ breaks 
the time-reversal invariance. For such a case, 
we find
$$
 H_{\hat{\bm{y}}}(q) =  \sinh q \ {\rm Im} (d) I_2 +{\rm Re}(d) \sigma_x -\cosh q\ {\rm Im}(d) \sigma_y
$$
with an energy dispersion, 
\begin{eqnarray*}
 E_q(k_1,k_2) &=& {\rm Im} (d)\sinh q \\
 & &\pm \sqrt{({\rm Re}(d))^2 + ({\rm Im}(d))^2\cosh^2 q},
\end{eqnarray*}
in which we have a symmetry $E_q(-k_1,-k_2)$$ = E_{-q}(k_1,k_2)$ (see Appendix A).  In this case, the Dirac cones at K and K' are tilted in the same direction (Fig.\ \ref{honeycomb_def_y}).  
The parameters for the effective low-energy Hamiltonian at K and K' points are 
summarized in Table \ref{tab1}.

\subsection{$\pi$-flux model}

\begin{figure}[h]
\includegraphics[scale=0.48]{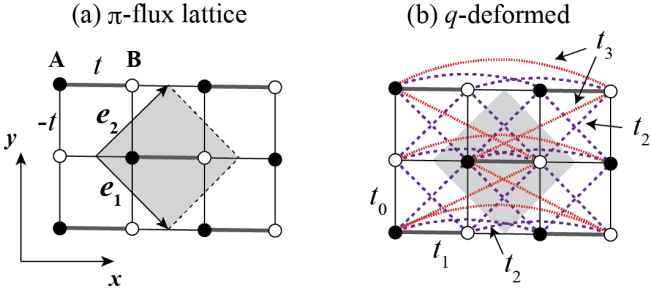}
\caption{
(a) The $\pi$-flux model on a square lattice, where we display 
a unit cell (shaded) 
and the 
primitive vectors  $\bm{e}_1=(1, -1)$ and $\bm{e}_2=(1,1)$ with 
the distance between the nearest-neighbor sites taken to be the unit of 
length. Thin lines represent the hopping amplitude $-t$ while the thick ones $t$.  
The filled (open) circles denote A(B) sub-lattice sites. 
(b) $q$-deformed $\pi$-flux model with $T_{\hat{\bm{x}}}(q)$. The hopping amplitudes are given by
$t_0 = -(t/2) (\cosh q+1)$ (thin lines), $t_1 = t\cosh q$ (thick), $t_2 = (t/2)\sinh q$ (dashed), and  $t_3 = -(t/2)(\cosh q -1)$ (dotted).  
The potential energies are modified uniformly to $-t\sinh q$.
\label{pi-flux}
}
\end{figure}

Another lattice model of interest is 
the $\pi$-flux model on the square lattice\cite{MH,HWK,KHMA} as depicted in Fig. \ref{pi-flux}(a). 
The Hamiltonian in real space is given by 
$$
 H_{\pi{\rm -flux}} =-t \sum_{x,y}\left[ (-1)^{x+y}c_{\bm r}^\dagger c_{\bm{r} -\bm{e}_x} +c_{\bm r}^\dagger c_{\bm{r}+\bm{e}_y}\right] + {\rm H.c.},
$$ 
where $\bm{r} = x\bm{e}_x + y\bm{e}_y$ with $x,y:$ integers.
For this model, we have \cite{KHMA,HFA}
$$
 d(\bm{k}) = \\
-t\left[-1+\exp(-\mi k_1)+\exp(-\mi k_2)+\exp(-\mi (k_1+k_2))\right],
$$
where $(k_1, k_2) = (\bm{k}\cdot\bm{e}_1, \bm{k}\cdot\bm{e}_2)$ with the primitive vectors for the $\pi$-flux model 
shown in Fig. \ref{pi-flux}(b), 
$(\bm{e}_1, \bm{e}_2) = (\bm{e}_x - \bm{e}_y, \bm{e}_x +  \bm{e}_y)$ with the nearest-neighbor distance of the square lattice taken as the unit of length.  
This model has two Dirac points (which we shall also call K and K') at $(k_x,k_y) = (0,\pi/2)$ and $(0,-\pi/2)$.  

\begin{figure}[h]
\includegraphics[scale=0.6]{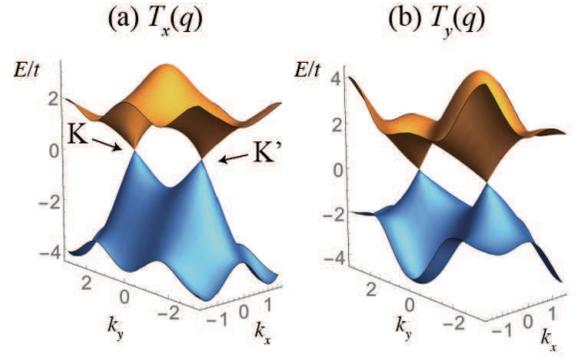}
\caption{(Color online)
Energy dispersions of the $\pi$-flux model deformed by (a) $T_{\hat{\bm{x}}}(q)$ and by (b) $T_{\hat{\bm{y}}}(q)$ with $q=0.5$. 
Dirac cones at K and K' points are tilted along $+k_y$ and $-k_y$ directions, respectively, for $T_{\hat{\bm{x}}}(q)$, while 
they are both tilted in the $-k_x$ direction for $T_{\hat{\bm{y}}}(q)$. The site energies are also 
modified uniformly to $-t\sinh q$.
\label{pi-flux_def}
}
\end{figure}

The parameters of the effective Hamiltonians at K and K' for $H_{\hat{\bm{x}}}(q)$ and  $H_{\hat{\bm{y}}}(q)$ are given in Table II.  
Again, we find that the Dirac cones are  tilted in opposite directions (Fig. \ref{pi-flux_def} (a)) by the deformation $T_{\hat{\bm{x}}}(q)$ 
due to the time-reversal invariance, while the cones are tilted in the same direction (Fig. \ref{pi-flux_def} (b)) for $H_{\hat{\bm{y}}}(q)$.

It is to be remarked  that the tilting direction in the $\pi$-flux model by the same operator $T_ {\hat{\bm{x}}}(q)$ (or $T_ {\hat{\bm{y}}}(q)$) is different from 
that in the honeycomb lattice. This is a clear demonstration of the fact that the tilting direction is actually determined not only by the choice of $\bm{\tau}$ but 
by the parameters $\bm{X}$ and $\bm{Y}$ in the effective Hamiltonian at each valley.

\begin{table}
\begin{tabular}{c|cccc}
 & $X_0$ & $Y_0$ & $\bm{X}$ & $\bm{Y}$\\
\hline \hline
$H_{\hat{\bm{x}}}(q)$ at K & $0$ &$\sinh q$& $(0,-1,0)$ & $(\cosh q,0,0)$ \\
$H_{\hat{\bm{x}}}(q)$ at K' & $0$ &$-\sinh q$& $(0,-1,0)$ & $(-\cosh q,0,0)$\\
$H_{\hat{\bm{y}}}(q)$ at K & $-\sinh q$ &$0$& $(0,-\cosh q,0)$ & $(1,0,0)$ \\
$H_{\hat{\bm{y}}}(q)$ at K' & $-\sinh q$ &$0$& $(0,-\cosh q,0)$ & $(-1,0,0)$\\
\hline
\end{tabular}
\caption{For the $\pi$-flux model the parameters $(X_0,Y_0,\bm{X},\bm{Y})$ normalized by $2ta/\hbar$ are given for the effective massless Dirac Hamiltonian $(-X_0+\bm{X}\cdot\bm{\sigma})p_x +(-Y_0+\bm{Y}\cdot\bm{\sigma})p_y$ at valleys K and K'. For a general case where $\bm{\tau}$ is given by $(\cos \theta ,\sin \theta,0)
$, we have $X_0=-\sinh q\sin \theta$, $Y_0=\sinh q \cos \theta$, $\bm{X}=(-\sinh^2\frac{q}{2}\sin2\theta,-\cosh^2\frac{q}{2}+\sinh^2\frac{q}{2}\cos2\theta,0)$, $\bm{Y}=(\cosh^2\frac{q}{2}+\sinh^2\frac{q}{2}\cos2\theta, \sinh^2\frac{q}{2}\sin2\theta,0)$ for valley K. The parameters for valley K' are obtained 
by reversing the sign of $Y_0$ and $\bm{Y}$.}
\end{table}


\section{Application to Fermion-vortex systems}

One of the advantages of the present deformation scheme is that it can be 
applied to systems without translational invariance. 
Let us then take an example in the 
fermion-vortex system, where the zero modes  are expected to accommodate  
fractionally charged states \cite{HCM,RMHC,CHJMPS}. For a fermion-vortex system, it has been shown that there exist $n$ zero-energy states localized around 
the vortex with a winding number $n$ \cite{JR}. For a conventional fermion-vortex system, the Dirac cones are vertical and therefore the lattice models considered in the
previous studies 
respect the conventional chiral symmetry. Then the zero-energy states are 
simply eigenstates of the chiral operator having their amplitudes only on one of the 
A(B) sub-lattices. Thus it is intriguing to see 
how the zero-energy states  would be 
modified for tilted Dirac fermions, so we apply the present deformation 
with the generalized chiral
symmetry preserved to vortex systems.  

\begin{figure}[h]
\includegraphics[scale=0.45]{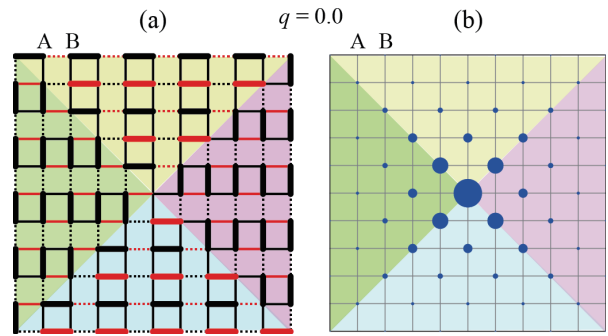}
\caption{(Color online)
(a) A vortex structure of dimer orders in the $\pi$-flux model. The amplitudes 
have $t+\delta t$ (thick solid lines), $t$ (thin solid lines) or 
$t-\delta t$ (thin dotted lines), where $\delta t (< t)$ denotes the strength of the dimer order. The sign of the hopping amplitudes is plus (minus) for the red (black) lines as in the undimerized 
$\pi$-flux model. 
(b) The zero-energy state around the vortex center obtained by the exact diagonalization of 
a finite system ($50 \times 50$) with $\delta t /t =0.6$ are shown.
The radius of circle is proportional to the amplitude at each site.
\label{pi-flux-vortex}
}
\end{figure}

As the starting Hamiltonian, we consider a vortex of a dimer order in the $\pi$-flux model as shown Fig. (\ref{pi-flux-vortex}) \cite{CHJMPS}. 
For that purpose, we introduce to 
the Hamiltonian $H_{\pi \rm{-flux}}$  four types of dimer orders, 
\begin{eqnarray*}
 H_{\rm dimer}^{\pm x} &=& \mp \delta t \sum_{\bm{r} \in A} (-1)^y (c_{\bm{r}+\bm{e}_y}^\dagger c_{\bm{r}}-
c_{\bm{r}-\bm{e}_y}^\dagger c_{\bm{r}})+ {\rm H.c.},\\
H_{\rm dimer}^{\pm y} &=& \pm \delta t \sum_{\bm{r} \in A} (-1)^y (c_{\bm{r}+\bm{e}_x}^\dagger c_{\bm{r}} +c_{\bm{r}-\bm{e}_x}^\dagger c_{\bm{r}})+{\rm H.c.}.
\end{eqnarray*}
We arrange these four orders (shaded in different colors in Fig.\ref{pi-flux-vortex}), i.e., $H_{\rm dimer}^{+x}$, $H_{\rm dimer}^{+y}$, $H_{\rm dimer}^{-x}$ and $H_{\rm dimer}^{-y}$ are introduced in the regions
$x>|y|$, $y>|x|$, $x<-|y|$ and $y<-|x|$, respectively.  Then we have a vortex 
at the center, which is assumed to be on B sub-lattice without a loss of 
generality.  
When the whole system is covered by one of the dimer orders, for example by $H_{\rm dimer}^{-x}$, the dimer order mixes the 
two Dirac points K and K', and the energies $E(\bm{k})$ is given by $E(\bm{k}) = \pm2[(\sin^2k_x +\cos^2k_y)+(\delta t)^2(\sin^2k_y)]^{1/2}$ with 
a gap $\pm 2\delta t$ at $(k_x,k_y) = (0,\pm \pi/2)$. The effective low-energy Hamiltonian can be expressed with a basis $\{$K$_A$, $i{\rm
K}_B$, K'$_A$, $i{\rm K'}_B$$\}$ as 
$$
 H_{\rm eff} = \left(\begin{array}{cccc}
 0 & \alpha k_- & 0 &-\Delta \\
 \alpha k_+ &0& \Delta &0\\
 0 & \Delta^* & 0 & \alpha k_+ \\
 -\Delta^* & 0 & \alpha k_-&0
 \end{array}\right),
$$
where $k_\pm = k_x\pm \mi k_y$, $\alpha = 2ta$, and $\Delta=|\Delta| e^{\mi \theta}$ with $|\Delta| = 2\delta t$. 
The phases of $\Delta$ for the orders $H_{\rm dimer}^{-x}$,  $H_{\rm dimer}^{+x}$ and $H_{\rm dimer}^{\pm y}$ can be assigned as $\theta = 0$, $\pi$, and $\pm \pi/2$,
respectively. The winding number $n$ of the present vortex is thus $n=-1$. It has been shown \cite{JR,HCM} that such a vortex has a zero-energy state residing only 
on the B sub-lattice. This can be verified numerically by diagonalizing the Hamiltonian for a finite system with the vortex (Fig. \ref{pi-flux-vortex}(a)), where  
the zero mode localized at the vortex indeed has its amplitude only on the B sub-lattice (Fig.\ \ref{pi-flux-vortex} (b)).

\begin{figure}[h]
\includegraphics[scale=0.55]{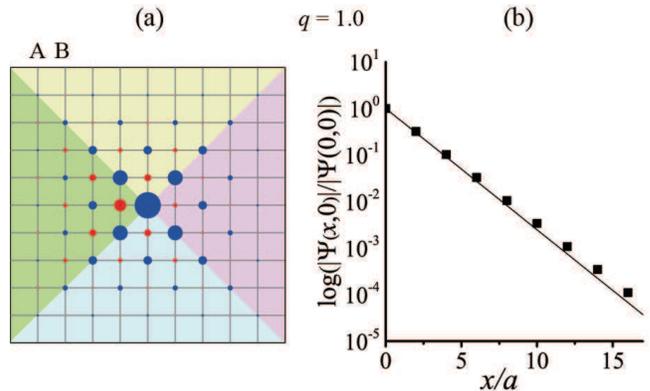}
\caption{(Color online)
(a) The zero-energy state of  the $q$-deformed Hamiltonian $H_{\rm vortex}^{(q)}$ 
with $q=1.0$. Amplitudes around the vortex obtained by the exact diagonalization of a finite system ($50 \times 50$) with $\delta t /t =0.6$ are shown. 
The radius of circle is proportional to the amplitude at each site.  
(b) The amplitudes of the zero-energy state along $x$-axis against 
the distance from the center of the vortex for $q=1$ (dots) 
are compared with $\exp(-(\delta t/t)(x/a)) = \exp(- 0.6(x/a))$(solid line). 
The amplitudes are normalized by the value at the center for $q=1$. 
\label{zero_wf}
}
\end{figure}

The generalization of such zero modes to the tilted Dirac fermions can be carried out by the present deformation. If we denote the original Hamiltonian 
with a vortex shown in Fig.\ \ref{pi-flux-vortex}(a) as $H_{\rm vortex}^{(0)}$, a $q$-deformed Hamiltonian $H_{\rm vortex}^{(q)}$ can be defined as
$$
 H_{\rm vortex}^{(q)} = T_{\hat{\bm{x}}}(q)^{-1}\  H_{\rm vortex}^{(0)}\ T_{\hat{\bm{x}}}(q)^{-1}
$$
with 
$$
 T_{\hat{\bm{x}}}(q) = I_N \otimes \exp\left(\frac{q}{2}\sigma_x\right).
$$
Here we consider a time-reversal invariant deformation.  
The zero-energy state $\psi_q$ for the $q$-deformed Hamiltonian turns out to 
exist, and its spatial profile is obtained as shown Fig.\ \ref{zero_wf}(a). 
We can immediately notice that the amplitudes reside not only on B sub-lattices but also on A.  
Note that the zero-energy state $\psi_q$ of the deformed Hamiltonian is related to that of the original Hamiltonian via $\psi_q = 
T_{\hat{\bm{x}}}(q) \psi_0$. Since the operation of $T_{\hat{\bm{x}}}$ simply  results in a modification of the wave function in each unit cell,
which is 
uniform over unit cells, the spatial behavior (decay, etc) of the wave function is little affected by the deformation. Specifically, 
the decay-rate of the zero-energy state is unaffected by the present deformation, implying the size of the vortex state is 
insensitive to the tilting of the Dirac dispersion. It is to be noted that 
the bulk gap at $E=0$ within a single domain  depends on $q$
as 
$\Delta E_q = \pm 2\delta t   [(1-\tanh q)/(1+\tanh q)]^{1/2}$, which goes to zero as $q \to \infty$.  The decay rate, which is independent of $q$,
therefore behaves differently from  
the bulk gap. 
We have actually confirmed this numerically by the fact that the amplitude of the wave function 
for $q=1$ (normalized by its value at the vortex center)  and that for $q=0$ are indistinguishable over several orders of magnitudes and 
the exponential decay of the 
wave functions 
is well-described by $\propto \exp(-(\delta t/t) r)$ with $r$ the distance from the center of the vortex as shown in Fig. \ref{zero_wf}(b). 

The present deformation for the fermion-vortex system clearly shows that the zero energy states obtained by Jackiw and Rossi \cite{JR}, which are 
the eigenstates of the conventional chiral operator, can be 
extended to tilted Dirac fermions as the eigenstates of the generalized chiral operator.

\section{Summary}
We have proposed an algebraic deformation of the Hamiltonian in which the generalized chiral symmetry is rigorously preserved. The
deformation can be applied to a wide variety of lattice models with/without the translational invariance and provides a unified theoretical framework for 
the general two-dimensional Dirac fermions with/without tilting. By applying the deformation  to 
 conventional Dirac fermions on lattice models, we have indeed  generated systematically 
the general tilted Dirac fermions on lattice models with 
the rigorous generalized chiral symmetry.
Throughout the deformation, the zero-energy state is preserved as the exact eigenstate of the generalized chiral operator, where 
its wave function is given by a 
simple transformation of that of the original Hamiltonian. 
Since the transformation is uniform over the system, the spatial profile of zero modes are insensitive to tilting the Dirac cones. 
With such a deformation, we have shown that the zero modes of the fermion-vortex system can be generalized to 
tilted Dirac fermions as the eigenstates of the generalized chiral operator. 
A possible application of the present deformation to, e.g., 
realistic lattice models for massless Dirac fermions in organic materials with four sites in a unit cell
\cite{KKS,KKSF,KSFG,KNTSK,GFMP} that have considerably tilted Dirac cones is an interesting future problem.

\begin{acknowledgments}

The work was supported in part by JPSJ KAKENHI grant numbers 
JP15K05218 (TK), 
JP16K13845 (YH)
and JP26247064. 

\end{acknowledgments}
\appendix

\section{Energy dispersion and Time-reversal symmetry}
The time-reversal operator $\Theta$ for spinless particles is given by the complex conjugation operator $K$.
The original Hamiltonian for a bipartite lattice (\ref{Ham_k}) is expressed as 
$$
 H = \sum_{\bm{k}} a_{\bm{k}}^\dagger d(\bm{k}) b_{\bm{k}}  + {\rm H.c.}
$$ 
Here $a_{\bm{k}}$($b_{\bm{k}}$) denotes the fermion operator  with wave vector $\bm{k}$ 
on the A(B) sub-lattice. The time-reversal operation thus yields
$$
 \Theta H \Theta^{-1} = \sum_{\bm{k}} a_{-\bm{k}}^\dagger d(\bm{k})^* b_{-\bm{k}} ,
$$
hence the time-reversal invariance $\Theta H \Theta^{-1}  = H$ in the original Hamiltonian 
implies $d(\bm{k})^* = d(-\bm{k})$. 

For the deformation with $T_{\hat{\bm{x}}}(q) = \exp(q\sigma_x/2)$, we have
\begin{eqnarray*}
 \Theta H_{\hat{\bm{x}}} (q) \Theta^{-1} &=& \Theta T_{\hat{\bm{x}}}^{-1}(q) \Theta^{-1} (\Theta  H \Theta^{-1}) \Theta  T_{\hat{\bm{x}}}^{-1}(q) \Theta^{-1}\\
 &=& T_{\hat{\bm{x}}}^{-1}(q) H   T_{\hat{\bm{x}}}^{-1}(q) = H_{\hat{\bm{x}}} (q),
\end{eqnarray*}
which suggests that the time-reversal invariance is retained throughout the deformation. Note that $T_{\hat{\bm{x}}}(q)$ is real, so that 
$\Theta T_{\hat{\bm{x}}}(q) \Theta^{-1} = T_{\hat{\bm{x}}}(q)$. For an  eigenstate $\psi_{\bm{k}}$ having an eigenvalue $E(\bm{k})$ 
with a wave number $\bm{k}$, we also have 
$$
 H_{\hat{\bm{x}}}(q) \psi_{-\bm{k}} = H_{\hat{\bm{x}}}(q) \Theta \psi_{\bm{k}} = \Theta H_{\hat{\bm{x}}}(q) \psi_{\bm{k}} = E_q(\bm{k}) \psi_{-\bm{k}},
$$
which implies a symmetry $E_q(\bm{k}) = E_q(-\bm{k})$ for the energy dispersion.

For the case of $H_{\hat{\bm{y}}}(q)$,  on the other hand, we have $\Theta T_{\hat{\bm{y}}}(q) \Theta^{-1} = T_{\hat{\bm{y}}}(-q)$, hence 
$$
  \Theta H_{\hat{\bm{y}}} (q) \Theta^{-1}  = H_{\hat{\bm{y}}}(-q),
$$
leading to a symmetry $E_{-q}(\bm{k}) = E_q(-\bm{k})$.
The time-reversal symmetry is therefore broken in this case, and the deformed Hamiltonian indeed has complex transfer integrals.

\section{Lattice models with flat bands}
\begin{figure}[h]
\includegraphics[scale=0.5]{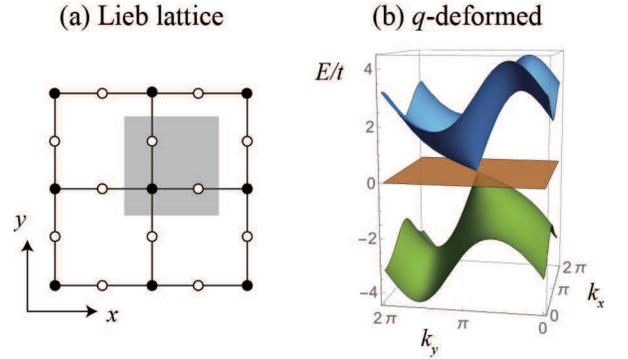}
\caption{(Color online)
(a) The Lieb lattice. The unit cell, which consists of three sites, is indicated by a shaded area. 
(b) Energy dispersion  for a $q$-deformed Lieb lattice, 
with $q=1$ here. 
\label{lieb}
}
\end{figure}

Here we show applications of the present deformation to the lattice models 
that accommodate flat bands on top of the massless Dirac fermions.
Let us first consider the Lieb lattice shown in Fig. \ref{lieb} (a), which 
is a prototypical flat-band model with a bipartite 
structure and hence respects the conventional chiral symmetry. 
A unit cell consists of three sites, where A  and B 
sub-lattices have different numbers of sites, with the 
difference giving the number of flat band(s) (here unity). 
The Hamiltonian in the momentum space is expressed as 
$$
 H= t \left(\begin{array}{ccc}
 0&1+\exp(\mi k_y)&1+\exp(\mi k_x)\\
 1+\exp(-\mi k_y)&0&0 \\
 1+\exp(-\mi k_x)&0&0 
 \end{array}\right),
$$
where the transfer integral between the nearest-neighbor sites is denoted by $t$ and we take the lattice constant as the unit of length. 
The energy eigenvalues are given by $E/t=0$ and $E/t=\pm \sqrt{|f(k_x)|^2+|f(k_y)|^2}$ with $f(k) = 1+\exp(\mi k)$, i.e., we have a flat band at 
$E=0$ which pierces a Dirac cone right at the Dirac point at 
$(k_x,k_y) = (\pi,\pi)$.

For this lattice Hamiltonian, we define the $q$-deformed Hamiltonian as 
$$
 H(q) = T(q)^{-1} H T(q)^{-1} , 
$$  
with 
\begin{eqnarray*}
 T(q) &=& \mmat{\exp(q\sigma_y/2)}{0}{0}{1}\\
 &=&\left(\begin{array}{ccc}
 \cosh(q/2)&-\mi \sinh(q/2)&0\\
 \mi \sinh(q/2)&\cosh(q/2)&0 \\
 0&0&1
 \end{array}\right).
\end{eqnarray*}
Then we can see that the deformed Hamiltonian $H(q)$ 
has an intact flat band at $E_q/t=0$ along with a deformed 
Dirac cone, 
\begin{eqnarray*}
 \lefteqn{E_q/t =   \sinh q \sin k_y} \\
& &  \pm \sqrt{|f(k_x)|^2\cosh q + [{\rm Re}\;f(k_y)]^2 + [{\rm Im}\;f(k_y)]^2\cosh^2q}.
\end{eqnarray*}
Namely, the deformed lattice model has a flat band 
that pierces a tilted Dirac fermion at the Dirac point, 
as depicted in Fig. \ref{lieb}(b).

\begin{figure}[h]
\includegraphics[scale=0.45]{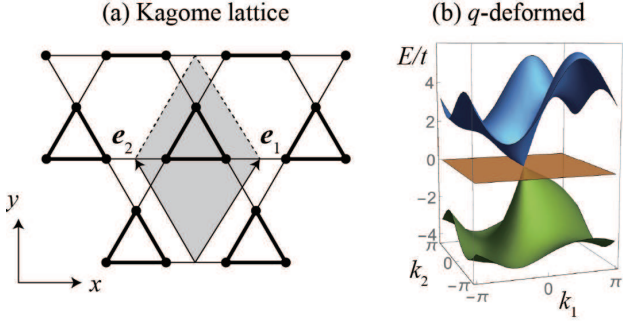}
\caption{(Color online)
(a) Kagome lattice, with thick (thin) lines denoting 
a transfer integral $-t (+t)$, for which the primitive vectors are 
$\bm{e}_1 = (1/2, \sqrt{3}/2)$,  $\bm{e}_2 = (-1/2,\sqrt{3}/2)$ 
with the lattice constant as a unit length.  
The unit cell, which consists of three sites, is indicated by a shaded area. 
(b) Energy dispersion for a $q$-deformed Kagome lattice, with $q=1$ here. 
\label{kagome}
}
\end{figure}

Another interesting example of lattice model with  a flat band coexisting 
with the massless Dirac fermion is Kagome lattice\cite{HM}.
When all of the hoppings are positive (negative),
 the flat band in the Kagome lattice is at the bottom (top) with doubled Dirac cones in the middle.  
If we modify the sign of the hoppings 
as depiected in Fig. \ref{kagome}(a), we can 
put the flat band as a middle one piercing a single Dirac point.
The Hamiltonian for this type of Kagome lattice is given in the momentum space as
$$
 H = -t \left(\begin{array}{ccc}
 0&1-\exp(\mi k_1)&1-\exp(\mi k_2)\\
 1-\exp(-\mi k_1)&0&1-\exp(\mi k_3) \\
 1-\exp(-\mi k_2)&1-\exp(-\mi k_3)&0 
 \end{array}\right).
$$
Here $k_i = \bm{k}\cdot \bm{e}_i \;(i=1,2)$ and $k_3 = k_2-k_1$ with $\bm{e}_1=(1/2,\sqrt{3}/2)$ and $\bm{e}_2=(-1/2,\sqrt{3}/2)$ being the 
primitive vectors of the Kagome lattice in units of the lattice constant (Fig.\ref{kagome}(a)).  
Energy dispersions then comprise the flat band at 
$E/t=0,$ along with $\pm 2\sqrt{\sin^2 ( k_1/2) +\sin^2( k_2/2) + \sin^2(k_3/2)}$ that has a massless Dirac cone located at $(k_x,k_y) = (0,0)$. 

For this model, we can apply the same deformation $H(q) = T(q)^{-1} H T(q)^{-1}$ as in the Lieb lattice.  
The eigenvalues of the deformed Hamiltonian 
then become
$$
 E_q/t =0, \quad \sinh q \sin k_1 \pm \sqrt{G(k_1,k_2)}
$$
with $G(k_1,k_2) = (1-\cos k_1)^2 + \sin^2k_1\cosh^2q+4(\sin^2\frac{k_2}{2} + \sin^2\frac{k_2-k_1}{2})\cosh q +2[\sin k_1-
\sin k_2+\sin(k_2-k_1)]\sinh q$. 
Namely, the flat band is again intact at $E=0$, while the Dirac fermion is 
tilted, as depicted in Fig.\ref{kagome}(b). 

According to the formalism in Ref.\ \onlinecite{HM}, the Hamiltonian with flat band(s) as in Kagome lattice is
expressed as a sum of (generically) non-orthogonal projections. Then, if the total dimension
of the projections is smaller than the dimension of the Hamiltonian itself,
the dimension of the null space, $d_N$, is nonzero ($d_N>0$). The null space 
corresponds to the 
zero-energy flat bands with a degeneracy $d_N$.
As for the Lieb lattice, which has the chiral symmetry with different numbers of sub-lattice sites, we can also apply such an approach \cite{HM} 
in 
describing the zero-energy flat band by considering $H^2$. 
Since the $q$-deformation introduced in the present work is a linear transformation, it is natural that 
the zero-energy flat bands remains unchanged.


\vfill
\end{document}